\documentclass[journal]{IEEEtran}
\usepackage{cite}
\usepackage{amsmath,amssymb,amsfonts}
\usepackage{algorithmic}
\usepackage[ruled,vlined,linesnumbered]{algorithm2e}
\SetAlCapNameFnt{\footnotesize}
\SetAlCapFnt{\footnotesize}
\SetAlgorithmName{Algorithm}{Algorithm}{List of Algorithms}
\usepackage{graphicx}
\usepackage{stmaryrd}
\usepackage{xcolor}
\usepackage{array}
\usepackage{commath}
\usepackage{footnote}
\usepackage{sidecap}
\usepackage{stfloats}
\usepackage{tabularx, boldline}
\usepackage{rotating,booktabs,multirow}
\usepackage{flexisym}
\usepackage{breqn}
\usepackage{url}
\usepackage{adjustbox}
\usepackage{makecell}

\usepackage{textcomp}
\usepackage{amsmath}
\usepackage{graphicx}
\usepackage{acronym}
\usepackage{tabularx, boldline}
\usepackage{rotating,booktabs,multirow}
\usepackage{enumerate}

\def\R{\mathbb{R}} 
\def\E{\mathbb{E}} 

\usepackage{authblk}
\def\BibTeX{{\rm B\kern-.05em{\sc i\kern-.025em b}\kern-.08em
    T\kern-.1667em\lower.7ex\hbox{E}\kern-.125emX}}

\usepackage[ruled]{algorithm2e}
\usepackage[hang,flushmargin]{footmisc}
\usepackage{lipsum}
\makeatletter
\newcommand{\algorithmfootnote}[2][\footnotesize]{%
  \let\old@algocf@finish\@algocf@finish
  \def\@algocf@finish{\old@algocf@finish
    \leavevmode\rlap{\begin{minipage}{\linewidth}
    #1#2
    \end{minipage}}%
  }%
}
\makeatother

\ifCLASSINFOpdf

\else

\fi

\begin{document}


	
\title{Privacy-Cost Management in Smart Meters with Mutual Information-Based  Reinforcement Learning}

\author{Mohammadhadi~Shateri,~\IEEEmembership{Member,~IEEE,} Francisco~Messina,
        Pablo~Piantanida,~\IEEEmembership{Senior Member,~IEEE,}
        Fabrice~Labeau,~\IEEEmembership{Senior Member,~IEEE}%

\thanks{M. Shateri and F. Labeau are with the Department of Electrical and Computer Engineering, McGill University, QC, Canada (e-mail: mohammadhadi.shateri@mail.mcgill.ca ; fabrice.labeau@mcgill.ca).}
\thanks{F. Messina is with the School of Engineering - Universidad de Buenos Aires, Buenos Aires, Argentina (e-mail: fmessina@fi.uba.ar).}
\thanks{P. Piantanida is with Laboratoire des Signaux et Syst\`emes (L2S), CentraleSup\'elec CNRS Universit\'e Paris-Saclay, 91190 Gif-sur-Yvette, France (e-mail: pablo.piantanida@centralesupelec.fr).}

}

\makeatletter
\def\ps@IEEEtitlepagestyle{
  \def\@oddfoot{\mycopyrightnotice}
  \def\@evenfoot{}
}
\def\mycopyrightnotice{
  {\footnotesize
  \begin{minipage}{\textwidth}
  Copyright (c) 20xx IEEE. Personal use of this material is permitted. However, permission to use this material for any other purposes must be obtained from the IEEE by sending a request to pubs-permissions@ieee.org.
  \end{minipage}
  }
}

\markboth{IEEE Internet of Things Journal}{}

\maketitle




%
\IEEEpeerreviewmaketitle

\begin{abstract}
	
The rapid development and expansion of the Internet of Things (IoT) paradigm has drastically increased the collection and exchange of data between sensors and systems, a phenomenon that raises serious privacy concerns in some domains. In particular, Smart Meters (SMs) share fine-grained electricity consumption of households with utility providers that can potentially violate users' privacy as sensitive information is leaked through the data. In order to enhance privacy, the electricity consumers can exploit the availability of physical resources such as a rechargeable battery (RB) to shape their power demand as dictated by a Privacy-Cost Management Unit (PCMU). In this paper, we present a novel method to learn the PCMU policy using Deep Reinforcement Learning (DRL). We adopt the mutual information (MI) between the user's demand load and the masked load seen by the power grid as a reliable and general privacy measure. Unlike previous studies, we model the whole temporal correlation in the data to learn the MI in its general form and use a neural network to estimate the MI-based reward signal to guide the PCMU learning process. This approach is combined with a model-free DRL algorithm known as the Deep Double Q-Learning (DDQL) method. The performance of the complete DDQL-MI algorithm is assessed empirically using an actual SMs dataset and compared with simpler privacy measures. Our results show significant improvements over state-of-the-art privacy-aware demand shaping methods. 

\end{abstract}

\begin{IEEEkeywords}
Internet of Things, Cyber-Physical System, Smart meters privacy, Mutual information, Deep reinforcement learning, Deep double Q-learning. 
\end{IEEEkeywords}

\section{Introduction}



With recent developments in computation, communication, and control technologies, the prevalent Internet of Things (IoT) facilitated the connection of billions of smart devices and sensors which can collect and exchange data in real-time\cite{mohammadi2018deep}. This has raised a lot of interest in Cyber-Physical Systems (CPS) for different applications such as Smart Grids (SGs) \cite{humayed2017cyber}. Smart Meters (SMs) are a key component of the so-called advanced metering infrastructure (AMI), which is a critical subsystem of SGs \cite{mohassel2014}. SMs are capable of measuring electricity consumption of users at a fine-grained level and share it with the Utility Provider (UP) in almost real-time. This large  amount of data provides immense opportunities for both customers and operators, leading to the emergence of the new field of SMs data analytics \cite{wang2019}. However, SMs data also contain sensitive information about users which could easily be inferred by malicious third-parties or attackers if no preventive measures are taken. For instance, an eavesdropper can apply non-intrusive load monitoring (NILM) methods and deep learning approaches to infer the user's presence at home \cite{giaconi2018privacy,9043691} and even the types of appliances being used at a specific time \cite{molina2010}. Therefore, the massive deployment and adoption of SMs necessitate the development of efficient privacy-aware strategies for real-time data sharing in order to keep the users' sensitive information private against potential attackers. It should be noted that the privacy issue regarding the SMs data sharing is distinct from SMs data security in terms of the legitimate users and the adversaries. Unlike the data security using encryption methods, in the SM data privacy any legitimate receiver of the data including utility provider can be considered at the same time as a potential malicious attacker. Therefore, the traditional encryption techniques would be ineffective in providing privacy against utility provider~\cite{giaconi2018privacy}. Recently, reinforcement learning has been used in designing models for preserving privacy in SM data sharing where physical resources at the users' end are used to prevent exposing consumers' privacy and behaviour to the UP~\cite{shateri9248831,li2018information,sun2017smart}. The main idea of reinforcement learning is based on the interaction between an agent (here a privacy-cost management unit) and the environment (which includes user and physical resources). More specifically, at each state imposed by the environment, the agent would take an action for the sake of receiving a maximum (future) reward and then being placed in the next state by the environment. These interactions can be used by the agent to learn the optimal policy which maximized the total reward~\cite{sutton2018reinforcement}. More details on the reinforcement learning and its formulation will be discussed in the Section~\ref{sec:prob_formulation}. Still a framework incorporates a generic privacy metric such as mutual information, is missing since it can add several challenges which requires some amendments in the traditional algorithms.

\subsection{Related work} \label{sec:related_work}

A substantial amount of studies on SMs privacy were conducted, which can be classified in two main families: (i) SMs data manipulation techniques \cite{efthymiou2010smart,sankar2013smart, yang2016evaluation, barbarosa2016, shateri2019deep, shateri2020a,9046241,gough2021preserving}; and (ii) user's demand load shaping approaches \cite{kalogridis2010privacy,yao2013privacy,tan2013increasing,gomez2014smart,zhang2016cost,giaconi2017smart,li2018information,erdemir2019privacy,giaconi2017optimal,sun2017smart, shateri9248831}. On the one hand, in the first family of methods, the consumers' load data are manipulated by a noisy transformation before sharing it to the UP. In this setting, there is a natural trade-off between the distortion or utility of the data and privacy guarantees. On the other hand, in the second family of privacy-aware techniques, the actual electricity consumption of the users, as seen by the grid (i.e., the grid load), are shaped using a combination of different physical resources such as Rechargeable Batteries (RBs), Electric Vehicles (EVs), Heating, Ventilation, and Air Conditioning (HVAC) units, and Renewable Energy Sources (RES). Note that, in this scenario, the consumer load is different than the grid load (e.g., the grid load can be higher than the consumer load provided that an RB is being charged by the user). The goal of these methods is to mask the consumer load but, at the same time, efficiently make use of the available resources considering their physical constraints and wear and tear, as well as a possibly time-varying electricity rate. In this framework, there is generally a trade-off between the overall electricity expenses and privacy guarantees.


In some recent studies, physical resources are employed to minimize the average relative difference between the grid load and a constant target load \cite{giaconi2017optimal,sun2017smart, shateri9248831}, i.e., to flatten the electricity consumption reported by the SMs. In \cite{shateri9248831}, following the formulation in \cite{sun2017smart}, the SMs privacy problem is cast as a Markov Decision Process (MDP) and a model-free (i.e., not assuming full knowledge of the environment dynamics of the MDP) Deep Reinforcement Learning (DRL) algorithm known as the Deep Double Q-Learning (DDQL) method is used to tackle the problem. Even though this framework has been shown to be useful in limiting the leakage of sensitive information, the effectiveness of the flatness-based privacy measure remains unclear.




A formal privacy measure from information theory known as the mutual information (MI), between the user's demand and reported grid load, was proposed in \cite{sankar2013smart} and has since been adopted in several works \cite{tan2013increasing,gomez2014smart,giaconi2017smart,li2018information,9249009}. In \cite{li2018information}, the SMs information-theoretic privacy problem is formulated as an MDP and the optimum policy is obtained numerically using dynamic programming, assuming full knowledge of the MDP. For a simplified scenario in which the demand load is assumed to be known and independent and identically distributed (i.i.d.), a single-letter expression for the average information leakage was characterized. However, these methods may not be directly applicable in practice since the MDP is generally not fully-known due to the unknown dynamics of the demand load. Furthermore, the electricity cost is not part of the formulation and thus, the cost-privacy trade-off is not taken into account.

\subsection{Contributions}

In this paper, we adopt a demand load shaping privacy-preserving strategy. As can be seen from the previous discussion, there is a clear gap between the DRL line of research and the information-theoretic work on this framework. As a matter of fact, the DRL methods rely on weak privacy measures, such as the flatness measure, which offer no statistical privacy guarantees (see Section \ref{sec:flatness} for further discussion). Whereas, information-theoretic approaches suggest the use of formal privacy measures, such as MI, with strong statistical privacy guarantees (see Section \ref{sec:mipm} for further details). The main goal of this paper, which further extends~\cite{shateri9248831}, is to develop a new DRL algorithm with the advantage of being model-free but also incorporating a strong information-theoretic privacy measure. Concretely, the main contribution is to incorporate the MI between the user's demand and grid load as a privacy measure within the DDQL algorithm framework. As will be shown, this introduces the challenge of estimating the privacy signal to feed into the DRL agent during the training phase, which is overcome by adding a new neural network (referred to as the H-network). Interestingly, based on the structure of this network, we can either use a general MI privacy measure or a simplified MI privacy measure based on a strong i.i.d. assumption. The latter case is an important benchmark as it allows us to quantify the importance of considering time correlation within the privacy measure. We then study the empirical information-leakage rate versus electricity cost trade-off on a real SMs dataset, and compare the MI-based and flatness-based privacy measures, showing the advantages of our approach. Finally, the performance of the DDQL-MI algorithm is assessed in two practical scenarios: (a) an attacker aiming to infer the actual consumer load; and (b) an attacker trying to infer the house occupancy status.

\emph{Organization of the paper.} The rest of the paper is organized as follows. In Section \ref{sec:prob_formulation}, we review the MDP formulation of the privacy-aware demand shaping problem. Then, in Section \ref{sec:privacy_measures}, we review two commonly used privacy measures and discuss how to use MI as a privacy measure in the MDP framework. We also discuss why MI is a superior privacy measure as compared with the flatness-based one. The novel DDQL-MI algorithm is then presented in Section \ref{sec:DQLMI}. The numerical performance of the DDQL-MI algorithm using actual SMs data is studied in Section \ref{sec:results}, where we show the impact of the privacy measure choice on the results. Some concluding remarks close the paper in Section \ref{sec:conclusion}.

\emph{Notation and conventions.} We use capital letter to denote random variables and lowercase to denote specific values. $P(y)$ is the probability distribution of random variable $Y$; $\E[\cdot]$ is expectation with respect to the joint distribution of all random variables involved, $\E[X|Y]$ is the conditional expectation of $X$ given $Y$, $\E_{\pi}[\cdot]$ is the expectation for a given agent's policy $\pi$ (see \cite{sutton2018reinforcement} for details); $H(Y)$ is the entropy of $Y$ and $I(X;Y) = H(X) - H(X|Y)$ is the MI between $X$ and $Y$, where $H(X|Y)$ is the conditional entropy of $X$ given $Y$ (see \cite{cover2006elements} for details).



\section{Problem Formulation} \label{sec:prob_formulation}

\subsection{Demand shaping using physical resources}

\begin{figure}[htbp]
	\centering
	\includegraphics[width=1\linewidth]{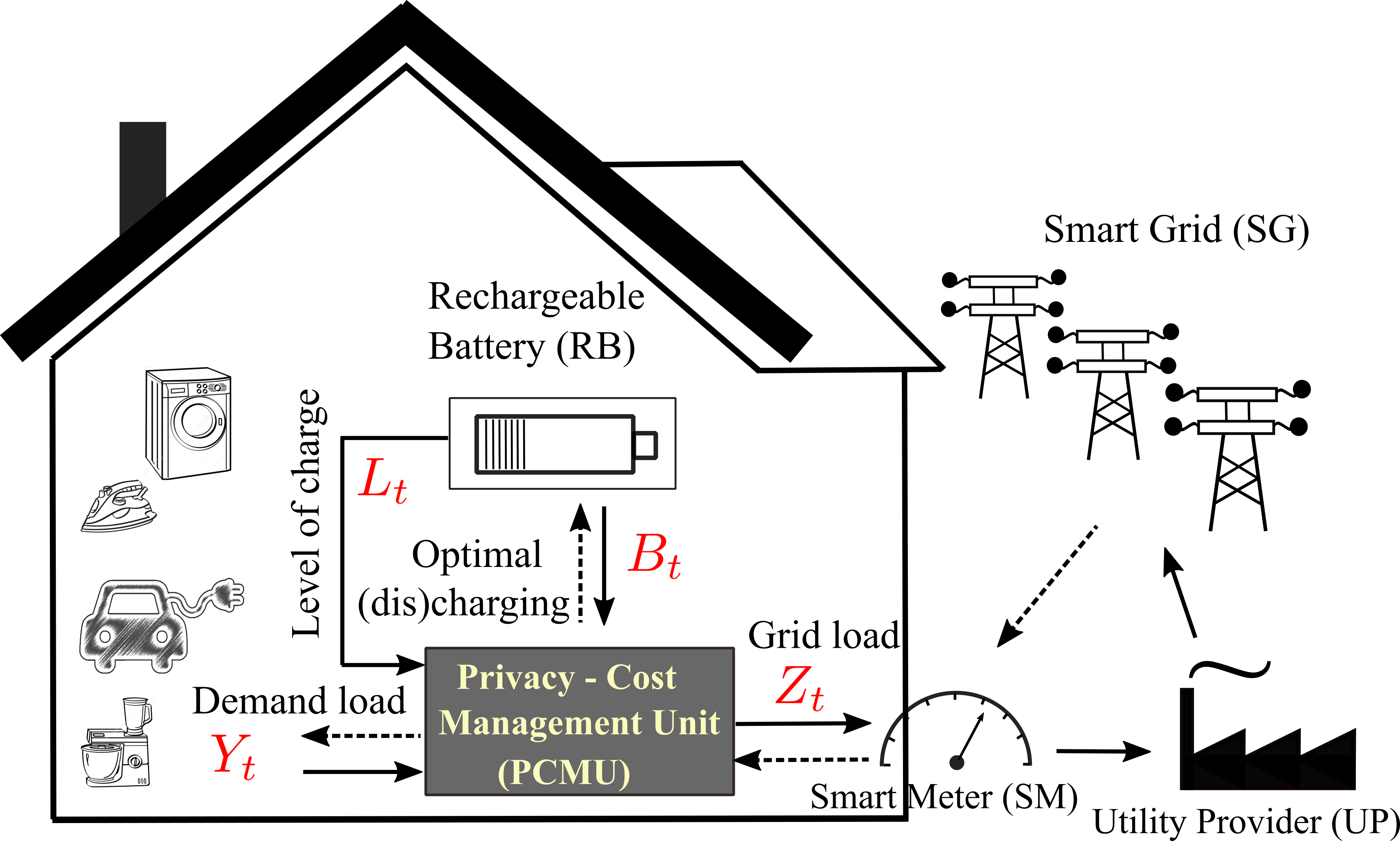}
	\caption{Privacy-aware demand shaping framework for smart meters based on a rechargeable battery~\cite{shateri9248831}.} 
	\label{RB-Based Model}
\end{figure}

Consider the smart metering system represented in Fig. \ref{RB-Based Model} where an intelligent agent, named as Privacy-Cost Management Unit (PCMU), is used to hide the household demand load using an RB while keeping the total electricity cost minimum. It should be noted that other physical resources can also be readily incorporated in our framework, as shown in other works \cite{sun2017smart,erdemir2019privacy}.

Let $Y_t$ be the random variable denoting the consumer's demand load/power, i.e., the total power demanded by the appliances at time $t$, and let $Z_t$ be the load received from the grid, and $L_t \in [0,1]$ the level of charge of battery (normalized by the capacity of the battery), where \mbox{$t \in \mathcal{T} =\{ 1, \ldots, T \}$}. Given the demand load $Y_t$ and the level of charge available in the battery $L_t$ at time $t$, the PCMU needs to determine the optimal charging/discharging rate of the battery $B_t$ to physically distort the actual demand load, so that the grid load $Z_t$, given by $Z_t = Y_t + B_t$ does not reveal information about the user's demand load $Y_t$. The goal of the PCMU is to limit the performance of a potential attacker trying to violate the user's privacy by inferring sensitive information (which could be either the actual demand load $Y_t$ or a correlated variable of interest to the attacker) from $Z_t$. To make this precise, we first introduce the MDP formulation of the problem.

\subsection{Markov Decision Process (MDP) model}\label{MDP_sec}

\begin{figure}[htbp]
	\centering
	\includegraphics[width=0.7\linewidth]{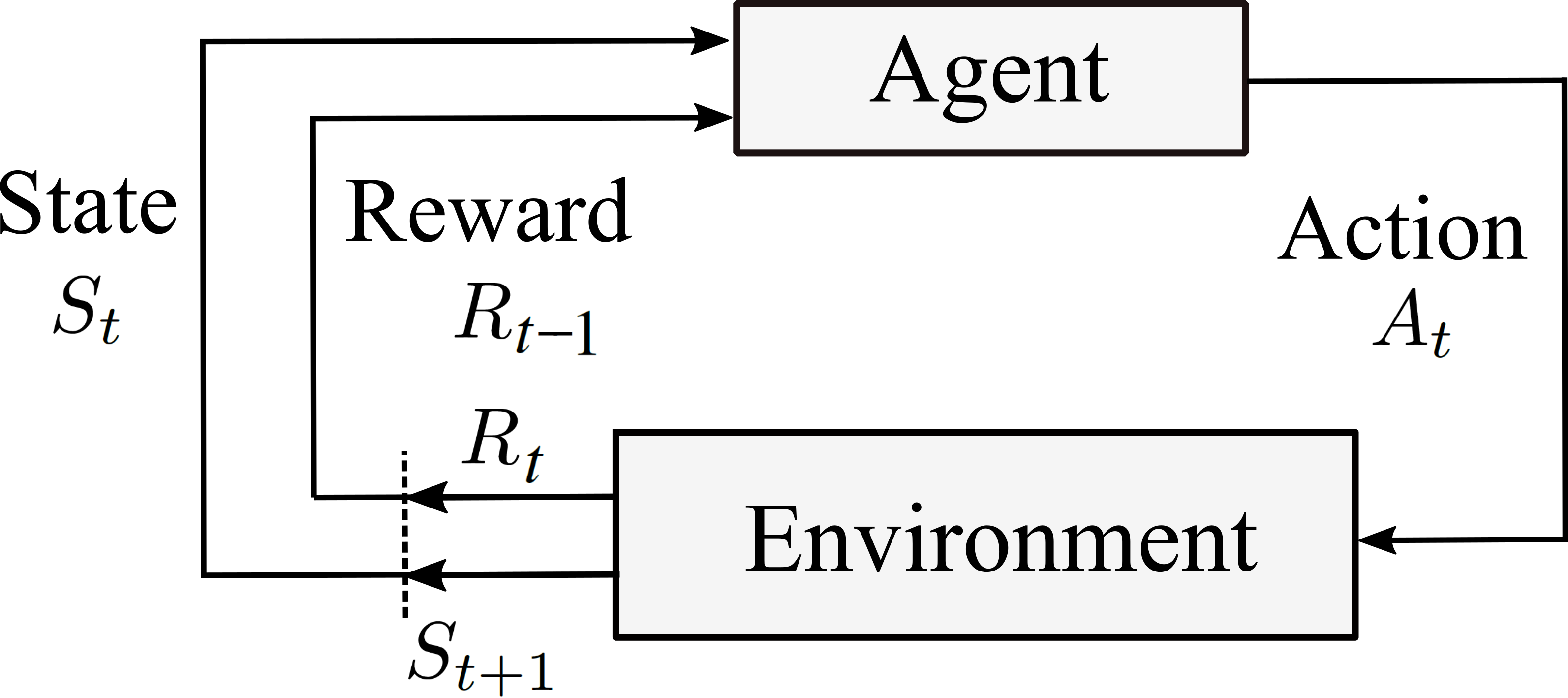}
	\caption{Agent-Environment interaction in an MDP \cite{sutton2018reinforcement}.}
	\label{Agent-Env-int}
\end{figure}

Following previous studies \cite{sun2017smart,li2018information,erdemir2019privacy}, the problem of finding the optimal policy for the PCMU is formulated as a Markov Decision Process (MDP) to capture the agent-environment interaction (see Fig. \ref{Agent-Env-int}). An MDP is determined \cite{sutton2018reinforcement} by the tuple $\left(\mathcal{S},\mathcal{A}(s),P(s_{t+1}|s_t,a_t),r(s_t,a_t),\gamma\right)$: 
\begin{itemize}
    \item State space $\mathcal{S}$, which determines all the possible states that the agent could be in;
    \item Action space $\mathcal{A}(s)$, which determines the feasible actions the agent can take at state $s \in \mathcal{S}$;
    \item Environment dynamics $P(s_{t+1}|s_t,a_t)$, which gives the probability of the state evolving to $S_{t+1} = s_{t+1}$ when the current state is $S_t = s_t$ and the agent takes the action $A_t = a_t$;
    \item Reward function $r(s_t,a_t)$, which is the immediate reward obtained due to taking action $A_t = a_t$ at state $S_t = s_t$;
    \item Discount factor $\gamma \in [0,1]$, which is the decay constant of future rewards and therefore determines their importance to the agent. In our setting, we assume a fixed finite horizon, i.e., $T<\infty$ is a constant. Therefore, the discount factor can be considered as $\gamma = 1$ without any convergence issues \cite{sutton2018reinforcement}. Moreover, as it will become apparent in Section \ref{sec:mipm}, this is the most natural choice for our problem. 
\end{itemize}

In general, the PCMU starts from an initial state $S_1$, and by following a policy $\pi(a|s) = P(a|s)$, takes the first action $A_1$. As a consequence of the action $A_1$, the PCMU receives the reward $R_1$ from its (artificial) environment and transitions to the state $S_2$ (see Fig.~\ref{Agent-Env-int}). Thereby, this gives rise to a trajectory of states, actions and rewards $\left[S_1,A_1,R_1,\ldots,S_T,A_T,R_T\right]$, which is referred to as an episode.

One general approach to find the optimum policy is via the action-value function $Q : \mathcal{S} \times \mathcal{A} \to \R$, which is defined as follows\cite{sutton2018reinforcement}:
\begin{equation} \label{eq:qfuncdef} Q(s,a) = \E_{\pi}\left[\sum_{t=1}^{T} \gamma^{t-1} r\left(S_t,A_t\right) \; \Bigg | \; S_1=s,A_1=a\right], 
\end{equation} 
where the subscript in the expectation emphasizes that it is a function of the policy $\pi$ chosen by the agent. The action-value function represents the expected cumulative reward received by the agent when it starts at some state $s \in \mathcal{S}$, takes an action $a \in \mathcal{A}(s)$, and then follows the policy $\pi$. Therefore, an optimal policy $\pi^*$ can be obtained by maximizing $Q(s,a)$ over all possible policies for all pairs $(s,a) \in \mathcal{S} \times \mathcal{A}(s)$. We use $Q^*$ to denote the optimal state-action value function, that is, $Q^*(s,a) = \max_{\pi} Q(s,a)$. It is well-known that, for any MDP, there always exists at least one optimal policy which is deterministic. Assuming uniqueness for the sake of simplicity of presentation, we can write the optimal policy simply as $\pi^*(a|s) = 1$ if $a = {\text{arg max}}_{a' \in \mathcal{A}(s)} \; Q^*(s,a')$ and $\pi^*(a|s) = 0$ otherwise.    



In our framework, the state at time $t$ is defined as \mbox{$S_t = [L_t, Y_t]^T$}, and the action is defined as the charging/discharging rate of the battery, i.e. $A_t = B_t$, where a positive $B_t$ indicates that the RB is charging and a negative $B_t$ means discharging of RB. For this MDP, the environment transition probability is $P(s_{t+1}|s_t, a_t)= P\left(l_{t+1}|l_{t},b_{t}\right) P\left(y_{t+1}|y_t\right)$, since we assume that the consumer's demand load is independent of the action and level of charge of battery. The factor $\small{P\left(l_{t+1}|l_{t},b_t\right)}$ is determined by the dynamics and physical constraints of the battery \cite{sun2017smart}:  

\begin{align} \label{eq:LOC-Dynamics} 
& L_{t+1} = L_{t} + \frac{B_t \, \Delta t \, \eta}{C},
\end{align}
where $B_t \in [b_{\min}, b_{\max}]$ ($b_{\min}$ and $b_{\max}$ are the minimum and maximum charging/discharging rate of the RB), \mbox{$L_{t} \in [l_{\min}, l_{\max}]$} ($l_{\min}$ and $l_{\max}$ are the minimum and maximum level of charge of the RB), $\Delta t$ is the load sampling rate, $\eta$ is the charging/discharging efficiency factor of the RB, and $C$ is the capacity of the RB. It should be noted that $P\left(y_{t+1}|y_t\right)$ is unknown and, in general, it is difficult to estimate accurately \cite{sun2017smart}. To get rid of this issue, the focus of this study is on model-free DRL algorithms which do not require full knowledge of the environment dynamics.

Finally, we need to define the reward function, which will guide the agent to learn an optimal policy. For our purposes, it should be an appropriate combination of the privacy leakage and the associated electricity cost. Following \cite{sun2017smart}, the reward function is inversely interpreted as a loss function: $\ell\left(s_t,a_t\right) = -r\left(s_t,a_t\right)$. Assuming a privacy leakage signal $f(s_t,a_t)$ and an electricity cost signal $g(s_t,a_t)$, which will be defined later, the one-step loss function can be defined as follows:
\begin{equation} \label{eq:CostFunction} \ell\left(s_t,a_t\right) = -r\left(s_t,a_t\right) = \lambda g(s_t,a_t) + (1-\lambda) f(s_t,a_t), \end{equation}
where $\lambda \in [0,1]$ controls the privacy-cost trade-off. Concretely, for $\lambda = 0$ the goal of the agent will be to minimize the expected cumulative privacy leakage signal, while for $\lambda = 1$ it will be to minimize the expected cumulative energy cost. As $\lambda$ is reduced, the PCMU should be able to provide the consumer a higher privacy level but at the expense of an increase in the energy cost. Studying this trade-off and its practical implications is of fundamental importance to properly design the PCMU. Notice that, using \eqref{eq:qfuncdef} and \eqref{eq:CostFunction}, we can decompose the action-value function as follows:
\begin{equation}\label{eq_action-value} Q(s,a) = \lambda Q_c(s,a) + (1-\lambda) Q_p(s,a), \end{equation}
where \mbox{$Q_c(s,a) = -\E_{\pi}[\sum_{t=1}^{T} \gamma^{t-1} g(S_t,A_t) \;| S_1=s,A_1=a]$} and \mbox{$Q_p(s,a) = -\E_{\pi}[\sum_{t=1}^{T} \gamma^{t-1} f(S_t,A_t) \;| S_1=s,A_1=a]$} are the cost and privacy action-value functions, respectively. Therefore, the objective of the PCMU, considering the aforementioned constraints on the capacity of the battery and its charging/discharging rate, is to find an optimal policy $\pi^{*}$ that maximizes the expected total reward in equation~\eqref{eq_action-value}. It should also be noted that the problem constraints are incorporated in the definitions of the state and action spaces. 

%

The total cost associated with this privacy-aware framework can be due to the electricity cost and also to the cost related to the battery wear and tear. Considering for simplicity that no energy can be sold to the grid by the users, the electricity cost at time $t$ can be computed as $C_t = \Delta t \; h_t  \; \left [ Z_{t} \right ]^{+}$ where $h_t$ is the price of purchasing $1$ kWh of energy from the grid at time $t$ and $[Z_t]^+ = \max(Z_t,0)$. Since $\left [ Z_{t} \right ]^{+} \leq \left|Z_t\right| \leq  Y_{t} + \left|B_t \right|,$ and $Y_{t}$ is not controlled by the PCMU, we consider the following electricity cost signal: 

\begin{equation} g(S_t,A_t) = \Delta t \; h_t  \; |B_t|. \end{equation}
Notice that \mbox{$C_t \le \Delta t \; h_t  \; (Y_t +|B_t|)$}, so this electricity cost signal effectively limits the actual electricity cost. In addition, this definition of $g(S_t,A_t)$ incidentally takes into account the battery wear and tear cost, since it grows as the battery use increases. The design of the privacy leakage signal is discussed in detail in the next section.

%

\section{Privacy Measures: Flatness and Mutual Information} \label{sec:privacy_measures}

In the following, two different privacy measures will be reviewed and discussed: the flatness privacy measure and the MI privacy measure.

\subsection{Flatness privacy measure} \label{sec:flatness} 

\begin{figure}[t]
	\centering
	\includegraphics[width=1\linewidth]{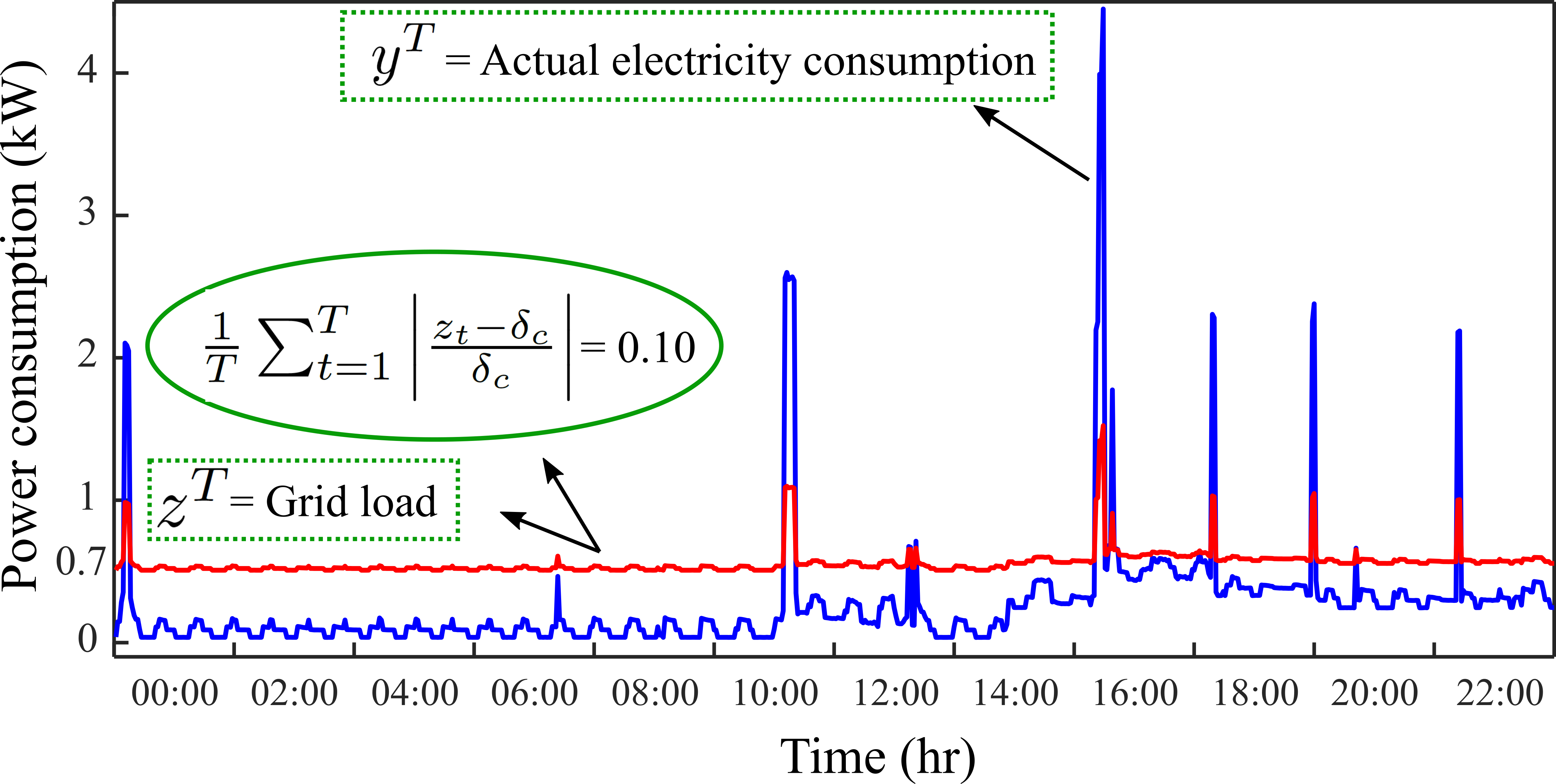}
	\caption{Demand and grid load when $z_t = \beta y_t + \delta_c$, where $\beta = 0.2$ and $\delta_c = 0.7$. Here we consider $T = 96$.} 
	\label{fig:flat_load}
\end{figure}

In \cite{sun2017smart}, the privacy leakage signal received by the Reinforcement Learning (RL) agent when at state $S_t$ and taking action $A_t$ is defined as
\begin{equation} \label{eq:PrivacyMeasureI} f\left(S_t,A_t\right) = \left |\frac{Z_t - \delta_c}{\delta_c} \right |, \end{equation}
where $\delta_c$ is a constant target level. This encourages the agent to take actions such that, on average, $Z_t$ is as close as possible to $\delta_c$. 
Although simple to compute and intuitively appealing, the problem with this quantity is that it does not capture the statistical dependence between $Y^T$ and $Z^T$ and therefore fails to be a satisfactory privacy measure. As an example to illustrate this fact, consider a PCMU with the following strategy: $Z_t = \beta Y_t + \delta_c$ for all $t \in \mathcal{T}$, where $\beta \in \R$ is a constant. In this case, the privacy action-value function, defined in \eqref{eq_action-value}, is 
\begin{equation} 
Q_p(s,a)  = - \E \left[\sum_{t=1}^{T} \gamma^{t-1} |\beta| \frac{|Y_t|}{|\delta_c|} \; \Bigg| \; s_1 = s, a_1 = a \right] \propto -|\beta|, 
\end{equation}
that is, $Q_p(s,a)$ is proportional to $-|\beta|$. If $\beta = 0$, we have $ Q_p(s,a) = 0$ as expected. In such scenario, $Z^T$ is constant and it does not provide any information to infer the actual value of $Y^T$. In other words, full privacy is achieved. However, for any $\beta \ne 0$, the variables $Y_t$ and $Z_t$ are maximally correlated (i.e., the correlation coefficient between $Z_t$ and $Y_t$ is either 1 or -1 depending on the sign of $\beta$) and the task of inferring $Y_t$ from $Z_t$ is trivial (assuming the attacker is able to estimate only two parameters). Therefore, all these cases can be considered equivalent from the privacy point of view, but $Q_p(s,a)$ can take any value in $(-\infty,0)$ as $|\beta|$ is modified. Thus, the flatness privacy measure can be completely misleading in some scenarios. This problem is illustrated in Fig. \ref{fig:flat_load}.

\subsection{Mutual information privacy measure} \label{sec:mipm} 

\subsubsection{General Case} \label{sec:migc}

A much stronger privacy measure, proposed in \cite{sankar2013smart} for the SMs privacy problem, and used since in several works \cite{tan2013increasing,gomez2014smart,giaconi2017smart,li2018information}, is the MI between the demand load and the grid load, which is defined as follows:
\begin{align} \label{eq:mi} I(Y^T;Z^T) & = \E \left[ \log \frac{p(Y^T,Z^T)}{p(Y^T) p(Z^T)} \right]. 
\end{align}
Intuitively, MI measures the degree of dependence between $Z^T$ and $Y^T$ and is zero if and only if $Z^T$ and $Y^T$ are statistically independent. It is also upper bounded by $H(Y^T)$, the entropy of $Y^T$, and equal to that value if and only if $Y^T$ is a deterministic function of $Z^T$ \cite{cover2006elements}. These standard properties of MI show why this quantity is satisfactory as a privacy measure. In fact, revisiting the previous example, in which $Z^T = \beta Y^T + \delta_c$, it readily follows that
\begin{equation} I(Y^T;Z^T) = \left\{
\begin{array}{lr}
0 & \text{if } \beta = 0, \\
H(Y^T) & \text{otherwise}. \\
\end{array}
\right.
\end{equation}
This means that full privacy is achieved only for the case $\beta = 0$  and other choices of $\beta$ lead to a maximal information leakage. Recall from \eqref{eq:CostFunction} that, in order to define the loss or the reward signal, we need to find the instantaneous (random) privacy leakage signal received by the agent when it is at state $s_t$ and takes action $a_t$. To do so, we first define the privacy action-value function as follows:
\begin{align} \label{eq:mi2} Q_p(s,a) &= - I(Y^T;Z^T) = \sum_{t=1}^T H(Y_t | Y^{t-1}, Z^T) - H(Y^T). \end{align}
It should be noted that $Q_p(s,a)$ is a function of the policy of the agent $\pi$ but not of the initial state and action. Notice that $H(Y^T)$ is a constant, independent of the PCMU strategy, so the second term in \eqref{eq:mi2} can be discarded. On the other hand, using the definition of conditional entropy and the law of total expectation, the term inside the summation of the first term can be written as follows:
\begin{align} \label{eq:EXP_priv} 
&H(Y_t|Y^{t-1},Z^T) = \E[\E[-\log P(Y_t|Y^{t-1},Z^T) | S_t , A_t]]. \end{align}
Notice that the inner conditional expectation is an explicit function of $S_t$ and $A_t$, as required. Therefore, we define the privacy leakage signal as follows: 
\begin{align} \label{eq:PrivacyMeasureII} f\left(s_t,a_t\right) = \E[ \log P(Y_t|Y^{t-1},Z^T) |S_t = s_t , A_t = a_t]. \end{align}
With this definition, the (negative) expected cumulative privacy leakage signal over an episode is equal to the MI in \eqref{eq:mi2} up to an additive constant. Notice that to recover the MI exactly we need to set $\gamma = 1$, i.e., do not discount future rewards. Therefore, we do not use discounting in this paper. As it was explained in Section \ref{MDP_sec}, this choice is in fact typical for finite horizon problems. 


Implementing the privacy leakage signal proposed in equation \eqref{eq:PrivacyMeasureII} poses three main challenges. First of all, to approximate this privacy signal, we need to estimate the unknown conditional distributions $P(y_t|y^{t-1},z^T)$ for each $t \in \mathcal{T}$. Secondly, although the expectation operation appearing in \eqref{eq:PrivacyMeasureII} can be estimated based on previous experiences of the agent using a Monte Carlo approach, it needs enough samples for each possible pairs of $(s_t,a_t)$ and so requires a huge buffer. Finally, unlike equation \eqref{eq:PrivacyMeasureI}, the approximation of this privacy measure is non-causal as it involves the whole sequence of the grid load $z^T$ at each $t$. In Section \ref{sec:DQLMI} we will discuss how to deal with these challenges.\\

\subsubsection{I.I.D. case} \label{sec:miiid}

In order to understand the role of the correlation in time of the time series $Y^T$ and $Z^T$, we also consider, as a benchmark, the case in which $Y^T$ is assumed to be independent and identically distributed (i.i.d.) and we model the transformation between $Y_t$ and $Z_t$ as memoryless but arbitrary, i.e., $Z_t = g(Y_t)$ where $g$ is a fixed random transformation. Although this assumption clearly does not hold in our MDP framework nor in practice, it is interesting to study this scenario to assess the advantage of taking into account the correlations across time for the privacy measure computation. Note that, in this case, $Z^T$ is also i.i.d. 
In such a case, it can be shown that:
\begin{equation} I(Y^T;Z^T) = \sum_{t=1}^T I(Y_t;Z_t)  = T [H(Y) - H(Y|Z)], \end{equation}
where we have omitted the time dependence since all the pairs $(Y_t,Z_t)$ are assumed to be i.i.d. and therefore all the terms are equal. Similarly as before, noting the relation
\begin{equation} H(Y|Z) = \E [\E[-\log P(Y|Z) | S, A]], \end{equation}
we can define the privacy signal simply as
\begin{equation} \label{eq:PrivacyMeasureIII}
f(s,a) = \E[\log P(Y|Z) | S = s, A = a]. 
\end{equation}

Note that, this i.i.d. case is not guaranteed to provide any control over the value of $I(Y^T;Z^T)$ in the general case. In fact, by using the standard properties of MI, it can be shown that\cite{cover2006elements}
\begin{equation} I(Y^T;Z^T) \ge \frac{1}{T} \sum_{t=1}^{T} I(Y_t;Z_t). \end{equation}
In summary, the i.i.d. assumption leads to using a lower bound of the MI as a privacy measure and, therefore, does not offer real privacy guarantees. 

\section{Methodology and Algorithm} \label{sec:DQLMI}

We first review classical RL  and DRL algorithms which were used in previous works on this topic. Then, we extend the DRL algorithm to accommodate the MI privacy measure.

\subsection{Review of CQL algorithm}

The classical Q-Learning (CQL) algorithm is a simple method to learn the optimal state-action value function $Q^*$ by updating the action-value of the experienced state-action pairs. The algorithm can be summarized by the update equation:
\begin{align} \label{eq:QLearning}
\Delta Q\left(S_t, A_t\right) &= \alpha \; \bigg[ r\left(S_t, A_t\right) + \gamma \underset{a\in \mathcal{A}\left(S_{t+1}\right)}{\text{max}} Q\left(S_{t+1},a\right)\nonumber\\& - Q\left(S_t, A_t\right)\bigg], \end{align}
%
where $\alpha$ is the step size parameter. Details on the training process and convergence properties of the CQL method can be found in \cite{sutton2018reinforcement}. The CQL algorithm was used for smart meter privacy in \cite{sun2017smart}. However, the main drawback of this method is that it needs to visit all the state-action pairs several times to provide a good approximation of $Q^*$. Therefore, for large MDPs with many states and actions, convergence is usually very slow. This is the case in our problem if the action and state spaces are discretized with high accuracy.

\subsection{DDQL-MI algorithm} \label{sec:ddqlmi_algorithm}

\begin{figure}[t]
	\centering
	\includegraphics[width=1\linewidth]{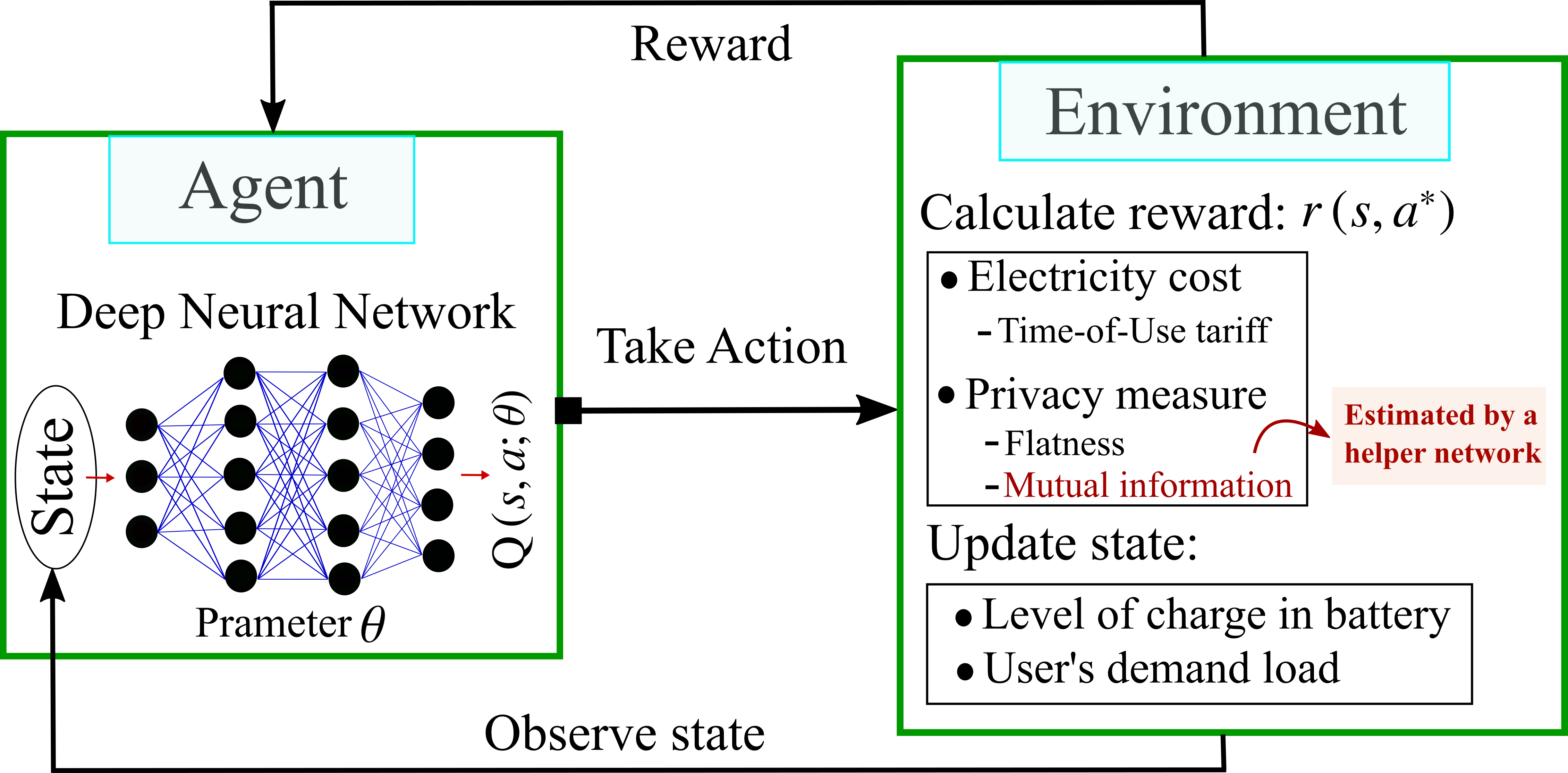}
	\caption{Schematic of the agent-environment interaction in the DQL framework where the Q-function is approximated with a deep neural network.}
	\label{Agent_Env}
\end{figure}

To solve the slow convergence problem of the CQL algorithm, the Q-function can be approximated by using a Deep Neural Network (DNN) to generalize between different states and actions. These new methods, where deep learning is used for approximating the Q-function, are called Deep Q-Learning (DQL) methods \cite{mnih2015human,franccois2018introduction}. A general diagram presenting the agent-environment interaction under the DQL paradigm and our context is shown in Fig. \ref{Agent_Env}.

The main idea of the DQL method is to approximate $Q^{*}(s,a)$ using a DNN called the Q-network. The Q-network takes the state $s\in\mathcal{S}$ at the input and generates $Q(s,a;\theta)$ at the output for all different actions $a\in\mathcal{A}(s)$, where $\theta$ are the parameters of the DNN. To define the objective function for this Q-network, we observe from \eqref{eq:QLearning} that convergence is obtained when the quantity in parenthesis is equal to zero. The term $r\left(s_t,a_t\right) + \gamma \; \text{max}_{a\in \mathcal{A}(s_{t+1})} Q\left(s_{t+1},a;\theta\right)$ can be interpreted as the target, while the term $Q\left(s_t,a_t;\theta\right)$ is the output of the Q-network. Thus, the mean squared error loss between target and output can be used as the loss function for training the Q-network. However, using the same network to compute the target and output often leads to instability \cite{mnih2015human}. To address this issue, the so-called Double Q-Learning (DDQL) algorithm was proposed in \cite{hasselt2010double} and extended to the deep learning setting in \cite{van2016deep}. In the DDQL algorithm, a second network called the target-network (with parameters $\theta^{\prime}$) is used to calculate the target term. The target-network parameters $\theta^{\prime}$ are periodically updated by simply copying the parameters from the Q-network. Thus, using the target-network, the objective function of the Q-network can be written as follows:
\begin{align} \label{eq:DDQL_loss}
\mathcal{L}_{QN}(\theta) =& \E\bigg[ \bigg(r(S_t,A_t) +
\gamma \underset{a\in \mathcal{A}(S_{t+1})}{\text{max}} Q(S_{t+1},a;\theta^{\prime})\nonumber\\
&-Q(S_t,A_t;\theta) \bigg)^2\bigg].
\end{align}
It should be noted that the expectation is approximated by a Monte Carlo approach based on batches of samples $(s_t,a_t,r(s_t,a_t),s_{t+1})$ selected randomly from a replay buffer \cite{franccois2018introduction}. The DDQL algorithm was applied for the first time in the SM privacy problem in \cite{shateri9248831} using the flatness privacy signal defined in \eqref{eq:PrivacyMeasureI}, showing clear improvements in performance and convergence speed with respect to the CQL algorithm. From now on, we will refer to this method as Model I.  

For the DDQL-MI method, we have to deal with the three challenges discussed in Section \ref{sec:migc} regarding the privacy leakage signal presented in equation \eqref{eq:PrivacyMeasureII}. To overcome the first challenge, a helper neural network named as H-network is included in the DD$\mathrm{Q}L$ to estimate the conditional probability distributions $p(y_t|y^{t-1},z^T)$, $t \in \mathcal{T}$. For the second challenge, instead of storing samples for each possible pairs of $(s_t,a_t)$ which leads to a huge storage and a very slow training, we approximate the privacy measure in equation \eqref{eq:PrivacyMeasureII} with its expected value over the joint distribution of $(s_t,a_t)$ , i.e., equation \eqref{eq:EXP_priv}. Then, for each episode, the pair $(y^T,z^T)$ is stored in a second replay buffer and, when required, samples are selected randomly to approximate equation \eqref{eq:EXP_priv}. Finally, since training the agent in DDQL is done offline, the third challenge is not an issue for its implementation. It should be emphasized at this point that, once a policy is learned, the agent will act according to it in a fully causal manner. In order to address the exploration$-$exploitation dilemma, we adopt the $\epsilon-$greedy method: with probability $1-\epsilon$ the action with maximum action-value $Q$ is selected (exploitation), while with probability $\epsilon$ a random action is selected (exploration). For more details, the reader is referred to \cite{sutton2018reinforcement}. The training of the DDQL-MI method is presented in Algorithm \ref{AlDDQL} below. It will be referred to as Model II in the following.

\begin{algorithm}
    \footnotesize
    \algsetup{linenosize=\tiny}
	\caption{DDQL-MI training algorithm.}
	\algorithmfootnote{Note: The copy step $k$ and training step $k^{\prime}$ are hyperparameters.}
	\label{AlDDQL}
	\begin{algorithmic}[1]
	    \STATE Initialize Q-network, target-network, and H-network. Initialize $l_{1}=0$.
		\FOR {number of training episodes}
		\STATE Set the initial state $s_1=[l_1,y_1]$.
		\FOR {$t=1,\ldots,T$}
		\STATE Observe the state $s_t=[l_t,y_t]$.
		\STATE Select a feasible action $a_t$ using the $\epsilon-$greedy algorithm.
		\STATE Calculate reward $r(s_t,a_t)$ from equation \eqref{eq:CostFunction}. 
		\STATE Update the next state $s_{t+1}$ based on \eqref{eq:LOC-Dynamics} and observing $y_{t+1}$.
		\STATE Import $(s_t,a_t,r(s_t,a_t),s_{t+1})$ into the replay buffer I.
		\STATE Every $k^{\prime}$ time-step, update the Q-network parameters using samples from the replay buffer I.
		\STATE Every $k$ time-step, update the target-network by copying from Q-network parameters.
		\STATE Every $k$ time-step, update the H-network and then estimate privacy measure in \eqref{eq:EXP_priv} for all $t \in \mathcal{T} =\{ 1, \ldots, T \}$ using samples from the replay buffer II. 
		\ENDFOR
		\STATE Import $(z^T,y^T)$ into the replay buffer II.
		\ENDFOR
	\end{algorithmic}
\end{algorithm}

It is worth to mention that the Q-learning algorithm has some shortcomings. First of all, for the sake of convergence, it needs a large amount of episodes to ensure all the state-action pairs are experienced multiple times. Secondly, the state and action sets need to be finite, while in practical applications it is not always the case~\cite{sutton2018reinforcement}. Finally, as we discussed in Section \ref{sec:privacy_measures}, a significant limitation of previous state-of-the-art Q-learning approaches to the smart meter privacy problem is the choice of the reward function, which does not capture a strong statistical notion of privacy. These shortcomings were considered in the design of Algorithm \ref{AlDDQL}. Concretely, regarding the first issue, the DQL algorithm is used instead of the CQL method in order to reduce the required number of training episodes. This point was discussed in detail in our previous work~\cite{shateri9248831}. In addition, the DQL algorithm is able to handle the infinite state-action space~\cite{franccois2018introduction}. Finally, the reward function design was revised (see Section \ref{sec:mipm}) to address the goal of this work, i.e. minimizing the information  leakage  about the users' electricity consumption pattern with a minimum increase in the electricity cost. As it was discussed, such a reward function raises several technical challenges. Thus, the training process of the Q-learning algorithm is carefully modified by adding an auxiliary network (H-network) to approximate the reward and help the agent in its learning. In Section~\ref{sec:results}, the convergence of this framework and the performance of the agent will be carefully examined to validate the proposed approach.

\section{Numerical Results and Discussion} \label{sec:results}

\subsection{Description of data set and parameters} \label{sec:dataset}

In this study, we use the public Electricity Consumption and Occupancy (ECO) dataset \cite{beckel2014eco}, which includes 1 Hz electricity usage measured by SMs along with the occupancy labels of five houses in Switzerland. The measurements sampling rate is chosen as $\Delta t=15$ min, and episodes with the length of a day are considered. In total, $2700$ samples (each a vector of length $T=96$) are used and split into training, validation, and test with ratio \mbox{70:10:20}, respectively. The training dataset was used to train the presented model while the values of the hyperparameters associated with the Q-network and the H-network were tuned using the validation dataset to achieve the best privacy-cost trade-off. After training and tuning, the performance of the model was evaluated based on the test dataset. For the flatness privacy measure, the desired constant load is set to $\delta_c = 0.7$ kW. The following values are considered for the parameters of the RB: $C  = 10$ kWh, $\eta = 1$, $b_{\max}= - b_{\min} = 4$ kW, $l_{\max}=1$ and $l_{\min}=0$. For the electricity cost calculations, since no time-of-use tariff was found online for Switzerland, the winter rates offered by Ontario/Canada is used, where the off-peak price is $\$0.101$ kWh during 19:00 to 7:00, the mid-peak price is $\$ 0.144$ kWh during 11:00 to 17:00, and the on-peak price is $\$0.208$ kWh during 7:00 to 11:00 and 17:00 to 19:00.

\subsection{Mutual information versus flat load as privacy } 

In this section, the results of applying Model I and Model II to the ECO dataset are presented. In both cases, a MultiLayer Perceptron (MLP) with two hidden layers, each including 64 neurons and Rectified Linear Unit (ReLU) as activation function, is used for both the Q-network and the target-network. The size of the experience replay memory is 10K tuples. The memory gets sampled to update the Q-network every 8 steps ($k^{\prime} = 8$), with minibatches of size 128, and a target-network copy step $k$ of 500 steps is selected. The RMSProp optimizer with a learning rate equal to 0.00025 is selected to train the network. For the H-network, the following two cases are considered. On the one hand, in the general MI case discussed in Section \ref{sec:migc}, a bidirectional RNN H-network with two hidden layers (each with 44 LSTM cells and hyperbolic tangent activation functions) is used.  The size of the second experience replay memory is 500 tuples and minibatches of size 64 are used. On the other hand, in the i.i.d. case where time dependency is ignored in calculating the MI, as discussed in Section \ref{sec:miiid}, the H-network is a feedforward neural network with two hidden layers (each with 64 neurons and ReLU activation functions). The size of the second experience replay memory is 10k tuples and minibatches of size 128 are used. In both cases, the cross-entropy loss  is used to train the network using the RMSProp optimizer with a learning rate equal to 0.001.

\begin{figure}[htbp]
	\centering
	\includegraphics[width=1\linewidth]{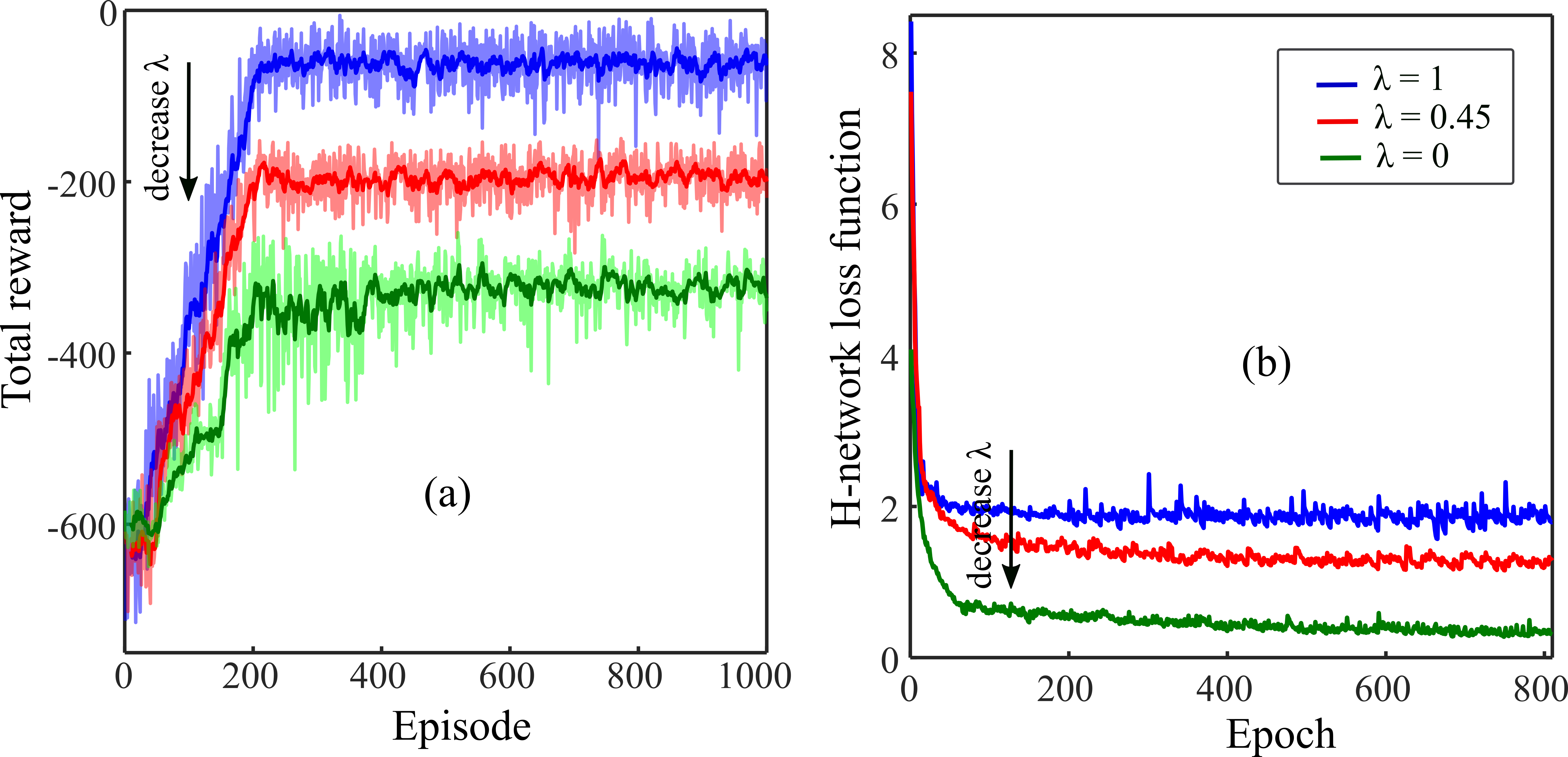}
	\caption{(a) Total episodic reward for the DDQL method. (b) The loss function of the H-network during training.}
	\label{rew_loss}
\end{figure}

Before presenting the results of both models, we need to show that the Algorithm 1 works. To this end, the results of the total episodic reward and the loss function of the H-network are presented in Fig. \ref{rew_loss} for different values of $\lambda$. It can be seen that both the reward function (determined by the H-network parameters) and the total episodic reward obtained by the agent converge in roughly 200 episodes. This suggests that the policy of the agent also converges.

\begin{figure}[htbp]
	\centering
	\includegraphics[width=0.7\linewidth]{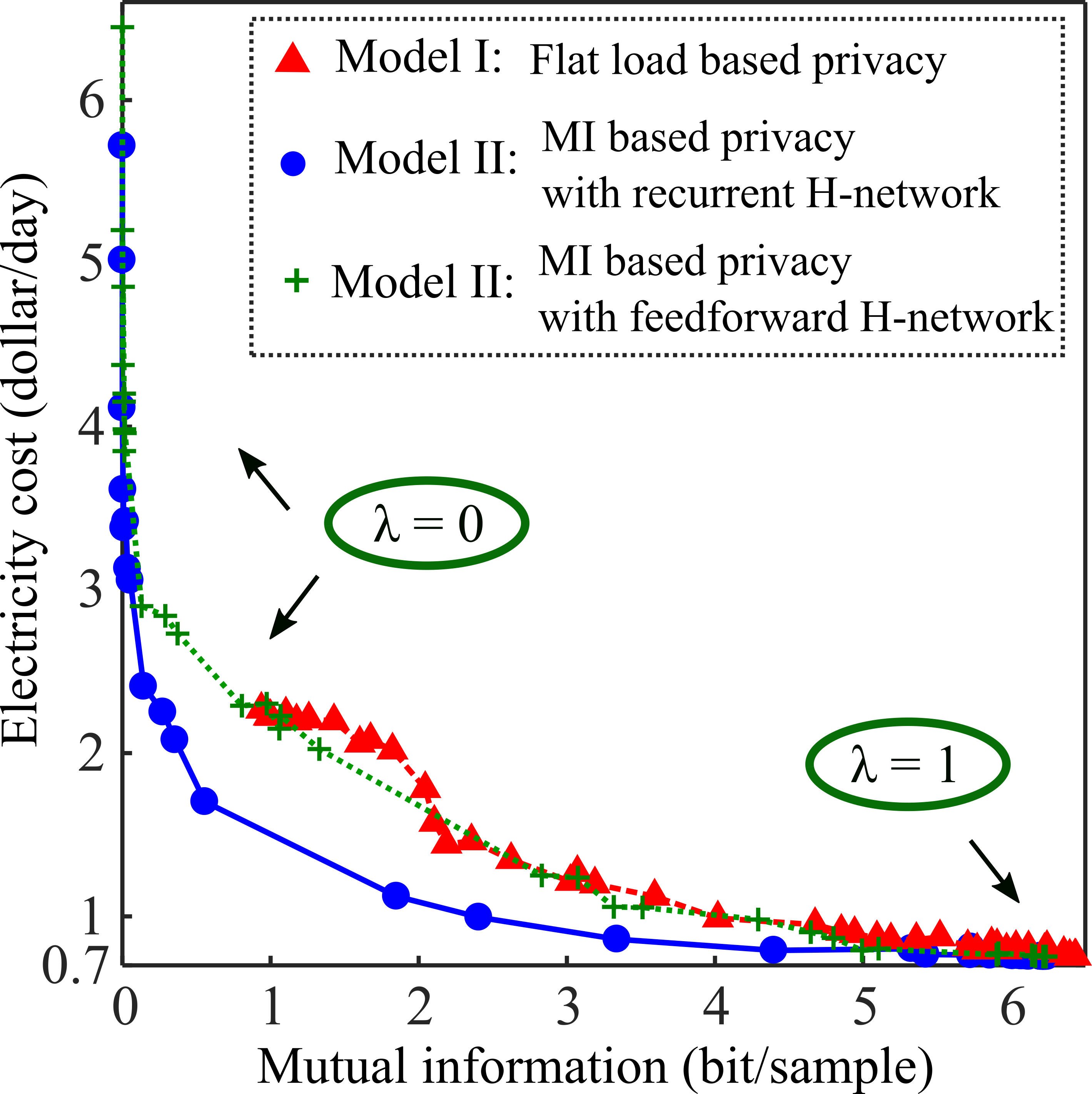}
	\caption{Electricity cost versus MI between demand load and grid load for Model I and Model II (for both a recurrent and a feedforward H-networks). }
	\label{MI_D}
\end{figure}

To compare the different methods, Fig. \ref{MI_D} presents the electricity cost versus the MI between demand load and grid load, calculated based on the Kraskov--St\"{o}gbauer--Grassberger (KSG) estimation method (with parameter 4). It should be noted that KSG method uses the k-nearest-neighbor distance of the points in dataset to estimate the underlying probability density needed to calculate MI. For more details about KSG method, readers are referred to \cite{kraskov2004estimating}. This figure clearly shows that the Model II outperforms Model I for two reasons. First, with the same electricity cost, Model II can provide a lower MI than Model I, which means that the statistical dependence between the sequences $y^T$ and $z^T$ is weaker for the former case. Second, unlike the Model I, Model II can provide MI up to very small values, thus offering the possibility of achieving practically arbitrary privacy levels. It should be noted that Model II is more computationally demanding than Model I in the training phase (due to the H-network required in Model II). However, in the operating phase, the computational cost is the same since only the Q-network is required for executing the learned policy. Another important analysis that can be made from Fig. \ref{MI_D} is the effect of using a recurrent H-network (general MI case) compared with a feedforward H-network (simplified i.i.d. case). As it was expected, the DDQL-MI using a recurrent H-network outperforms the one with a feedforward H-network, which can be seen from the gap of the curves in the figure. It should be added that, for values of $\lambda$ close to 1, since the privacy term has a very small weight in the loss function (see \eqref{eq:CostFunction}), all cases provide similar results.


\begin{figure}[htbp]
	\centering
	\includegraphics[width=1\linewidth]{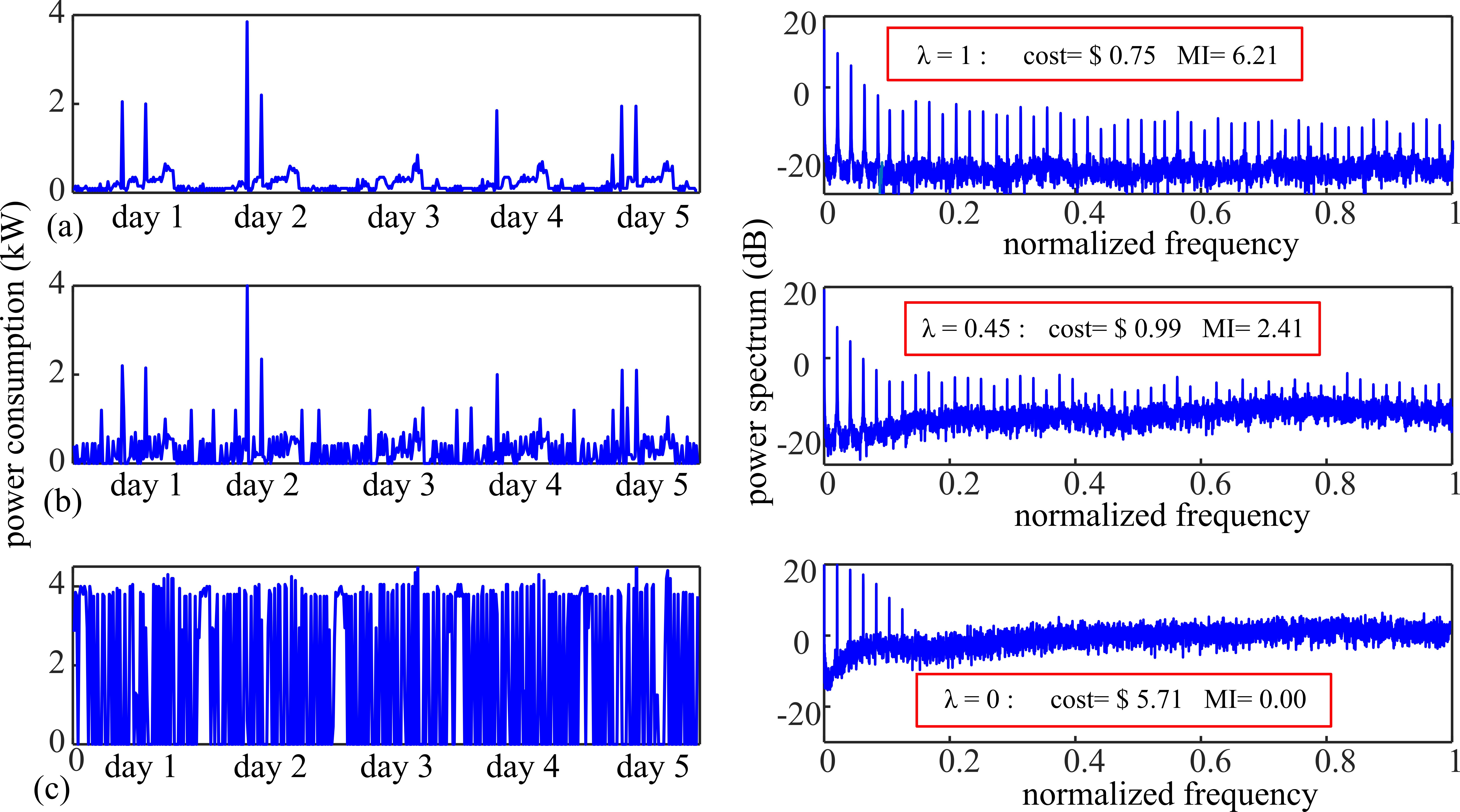}
	\caption{Examples of the grid load for different values of $\lambda$ along with its PSD estimated using the Welch’s method.}
	\label{visual_psd}
\end{figure}

In addition, for the model II with recurrent H-network, examples of the grid load signal for different values of $\lambda$ (along the trade-off curve in Fig. \ref{MI_D}) with its Power Spectrum Density (PSD) estimated using the Welch’s method~\cite{stoica2005spectral} are presented in Fig. \ref{visual_psd}. As it can be seen from this figure, for the middle values of $\lambda$, e.g. Fig. \ref{visual_psd}(b), the grid load signal looks like the actual demand load in Fig. \ref{visual_psd} (a) but is shaped to be a little bit noisier which would be for the sake of privacy. On the other hand, for $\lambda = 0$ the grid load looks very noisy in a way that completely hides the pattern of the actual demand load. This is reflected on the PSD in which the harmonics are hided more as we moves toward more private region. It should be noted that although the grid load signal for the full privacy case, i.e. $\lambda = 0$, would increase the electricity cost greatly as was expected, it could be of interest for the UP in terms of peak shaving.

\subsection{Deep double Q-learning versus attacker } \label{sec:DQLAttacker}

\begin{figure}[htbp]
	\centering
	\includegraphics[width=0.99\linewidth]{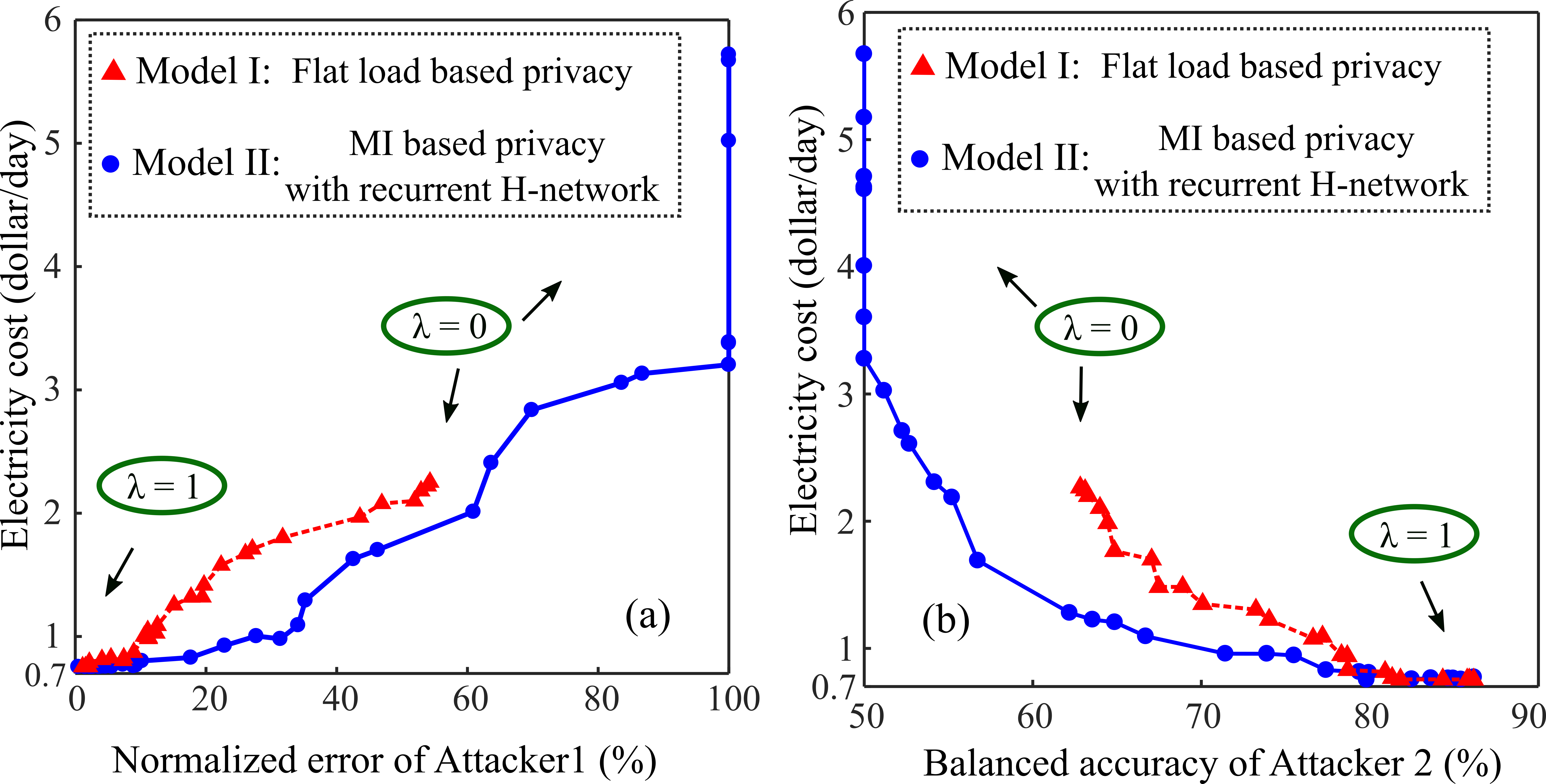}
	\caption{(a) Electricity cost versus normalized error of attacker inferring actual demand load and (b) Electricity cost versus balanced accuracy of attacker inferring occupancy status of household.}
	\label{attacker1_2}
\end{figure}

In this section, we evaluate the performance of Model I and Model II (using the general MI, i.e., the recurrent H-network) in limiting an attacker trying to infer sensitive information about the user. To this end, two practical scenarios are studied. In the first scenario, an attacker using a neural network with three hidden layers (each with 32 neurons and ReLU activation functions) uses the grid load sequence $Z^T$ to infer the user's demand load $Y^T$. In the second scenario, an attacker using a neural network with two hidden layers (each with 44 neurons and ReLU activation functions) uses the sequences of grid load $Z^T$ to infer the occupancy status of households. Both attackers are trained using the RMSProp optimizer with a learning rate equal to $0.001$. The performance of the first and second attacker versus the electricity cost is presented in Fig. \ref{attacker1_2}. From this figure, it can be seen that Model II is more effective in limiting the attackers since, for a given electricity cost, the inference performance metrics are worse in both cases (the exception, again, occurs in the regime $\lambda \approx 1$ where no privacy guarantees can be expected). Besides, when $\lambda \approx 0$, both attackers perform as expected when $Y^T$ and $Z^T$ are independent random vectors for Model II but not for Model I. This full privacy regime can be obtained at the expense of increasing the electricity cost. For example, looking at Fig. \ref{attacker1_2}(b), it can be seen that by increasing the electricity cost to more than $\$3/$day the attacker acts like random guessing in inferring the occupancy status of the dwelling. Note that this amounts to more than four times the normal electricity cost without privacy considerations.


\section{Summary and Concluding Remarks} \label{sec:conclusion}

In this work, we study a privacy-aware SM framework that uses an RB to hide the actual power consumption of a household. Following the literature, the problem of finding the optimum battery charging/discharging policy for minimizing information leakage with minimum additional electricity cost, is formulated as an MDP. This MDP is tackled using a model-free DRL approach, known as the DDQL algorithm. We propose to include the MI (between the actual power consumption and the masked grid load) as a strong privacy measure in the DDQL framework by using an H-network to estimate the required privacy leakage signal for training the agent. To evaluate the benefits of the proposed algorithm, the results are compared with the case where flatness is used as the privacy measure. The privacy-cost trade-off and the performance of two different attackers (attempting to infer sensitive information) are empirically obtained based on SM data, showing clear advantages of the new proposed method over the state-of-the-art on the topic. In addition, an i.i.d. scenario is considered as a benchmark to show the impact of the correlations across time in the privacy measure computation. It is shown that, by exploiting the time dependence, there is a consistent gain in the achieved privacy level for a given electricity cost. Although training our MI-based model is computationally more expensive than the others, its operating computational cost is equal if the structure of the Q-network is the same. Therefore, we conclude that the general DDQL-MI algorithm is able to better exploit an RB for privacy purposes using the same operating resources.  

To wrap up the paper, we briefly comment on two possible extensions of this work. First, it would be interesting to study different MDP formulations, where the definition of the state is wisely augmented to enhance the state observability, and analyze the performance gains that can be obtained. Second, a multi-user/multi-resource extension of this work also seems like a promising and challenging research avenue, where cooperation between different users is required.




\section*{Acknowledgment}
This work was supported by Hydro-Quebec, the Natural Sciences and Engineering Research Council of Canada, and McGill University in the framework of the NSERC/Hydro-Quebec Industrial Research Chair in Interactive Information Infrastructure for the Power Grid (IRCPJ406021-14). This project has received funding from the European Union’s Horizon 2020 research and innovation programme under the Marie Skłodowska-Curie grant agreement No 792464.

\bibliographystyle{ieeetr}
\bibliography{main}

\vspace{-6mm}

\begin{IEEEbiography}[{\includegraphics[width=1in,height=2in,clip,keepaspectratio]{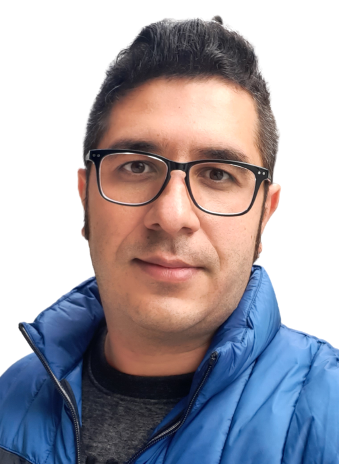}}]{Mohammadhadi Shateri}
(Member, IEEE) received the B.Sc. degree (with honors) in electrical engineering from the Amirkabir University of Technology, Tehran, Iran in 2012, the M.Sc. degree (with honors) in electrical engineering from the University of Manitoba, Winnipeg, Canada in 2017, and the Ph.D. in electrical engineering from McGill University, Montreal, Canada in 2021. His research interests include machine learning, deep learning, and reinforcement learning with application to data analytics.  

\end{IEEEbiography}

\begin{IEEEbiography}[{\includegraphics[width=1in,height=2in,clip,keepaspectratio]{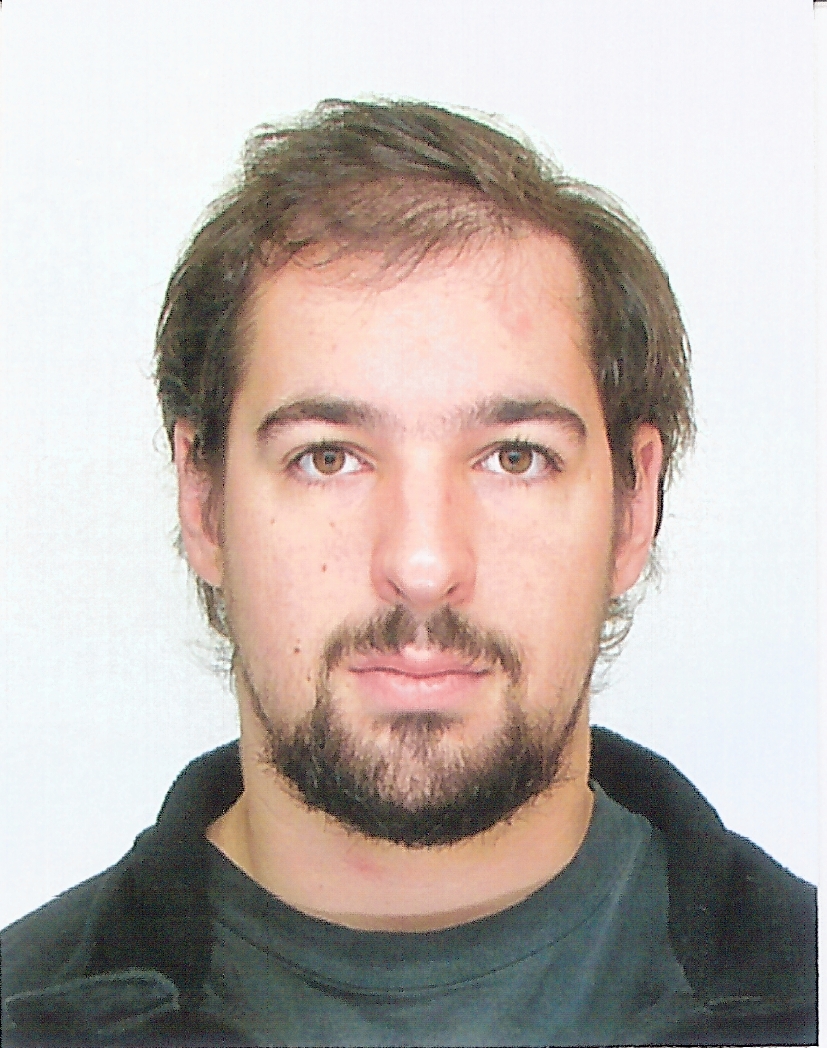}}]{Francisco Messina} received the M.Sc. and Ph.D. (Summa Cum Laude) degrees in electrical engineering from the University of Buenos Aires, Buenos Aires, Argentina, in 2014 and 2018, respectively. He was a Postdoctoral Fellow at McGill University, Montreal, Canada, between 2018 and 2020. Currently, he is a researcher at the University of Buenos Aires, Buenos Aires, Argentina. His research interests include signal processing and machine learning with a focus on their applications to smart grids. He has served as a reviewer for several IEEE conferences and journals.\vspace{-6mm}
\end{IEEEbiography}

\begin{IEEEbiography}[{\includegraphics[width=1in,height=2in,clip,keepaspectratio]{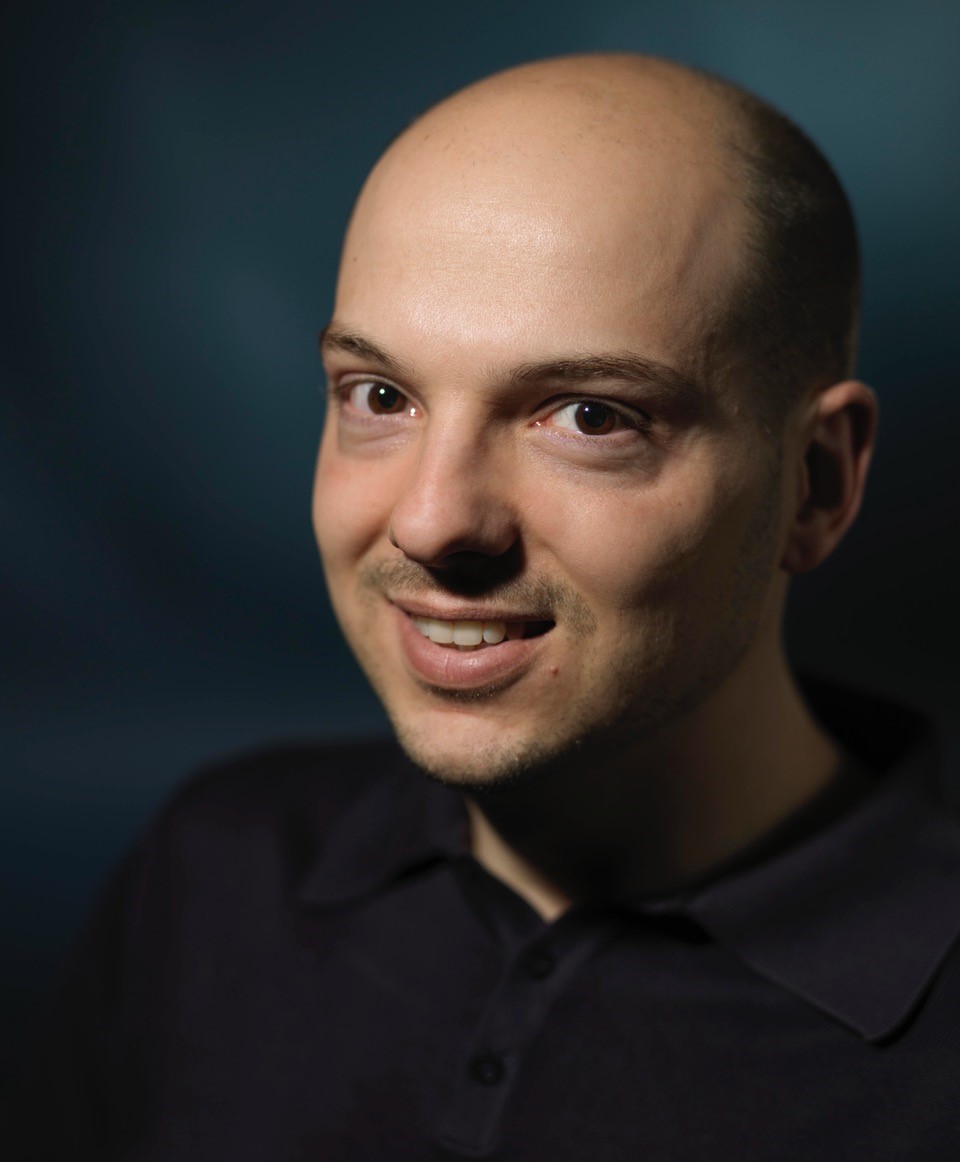}}]{Pablo Piantanida} (Senior Member, IEEE) received the  B.Sc.  degree in electrical engineering and the M.Sc. degree from the University of Buenos Aires, Argentina,   in   2003,   and the   Ph.D.   degree   from Université Paris-Sud, Orsay, France, in 2007. He is currently Full Professor with the Laboratoire des Signaux et Systèmes (L2S), CentraleSupélec together with CNRS and Université Paris-Saclay. He is also an associate member of Comète – Inria research team (Lix - Ecole Polytechnique). His research interests include information theory, machine learning, security of learning systems and the secure processing of information and applications to computer vision, health, natural language processing, among others. He has served as the General Co-Chair for the 2019 IEEE  International  Symposium on  Information  Theory  (ISIT).  He served as an Associate  Editor for the  IEEE  TRANSACTIONS  ON  INFORMATION FORENSICS AND SECURITY and Editorial Board of Section "Information Theory, Probability and Statistics" for Entropy. He is member of the IEEE Information Theory Society Conference Committee. 
\end{IEEEbiography}

\begin{IEEEbiography}[{\includegraphics[width=1in,height=2in,clip,keepaspectratio]{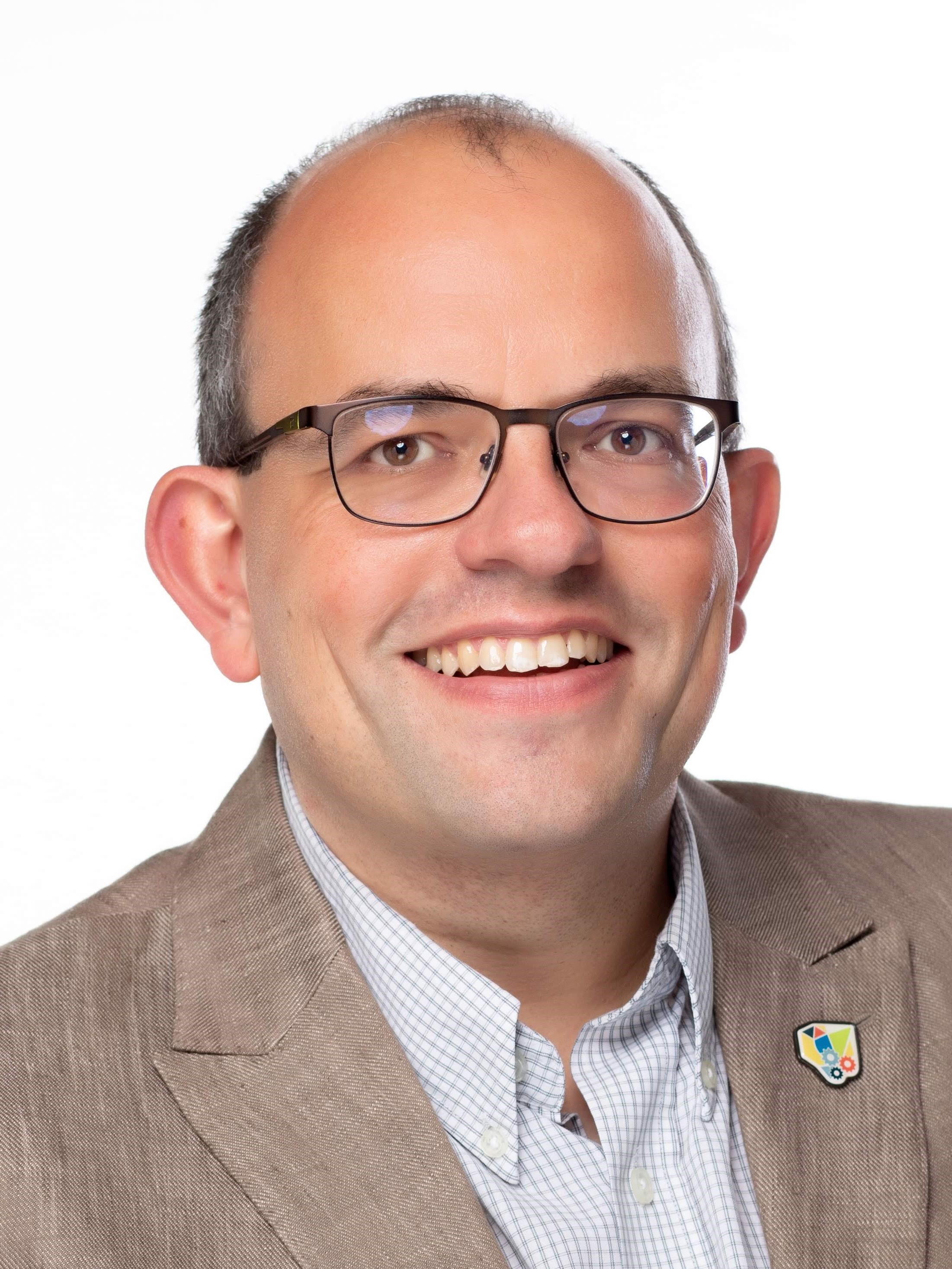}}]{Fabrice Labeau} is the Deputy Provost (Student Life and Learning) at McGill University, where he also holds the NSERC/Hydro-Québec Industrial Research Chair in Interactive Information Infrastructure for the Power Grid. His research interests are in applications of signal processing. He has (co-)authored more than 200 papers in refereed journals and conference proceedings in these areas. He is the Director of Operations of STARaCom, an interuniversity research center grouping 50 professors and 500 researchers from 10 universities in the province of Quebec, Canada. He is Past President of the IEEE Sensors Council, former President (2014-2015) of the IEEE Vehicular Technology Society, and a former chair of the Montreal IEEE Section. He was a recipient in 2015 and 2017 of the McGill University Equity and Community Building Award (team category), of the 2008 and 2016 Outstanding Service Award from the IEEE Vehicular Technology Society and of the 2017 W.S. Read Outstanding Service Award from IEEE Canada. He was recognized in 2018 "Ambassadeur Accrédité" for the Montreal Convention Center. He is a "champion" for Engineers Canada's 30 by 30 initiative.\vspace{-6mm}
\end{IEEEbiography}

\end{document}